%% file: main.tex
\documentclass[lettersize,journal]{IEEEtran}

\usepackage{url}
\usepackage{xspace}
\usepackage{booktabs}
\usepackage{makecell}
\usepackage{hyperref}
\usepackage{graphicx}
\usepackage{colortbl}
\usepackage{multirow}
\usepackage{subfigure}
\usepackage{color, xcolor}
\usepackage{amsthm,amsmath,amsfonts}
\usepackage{pifont}
\usepackage{enumitem}
\usepackage{rotating}
\usepackage[ruled,vlined,linesnumbered]{algorithm2e}
\usepackage{wrapfig}

\usepackage{xspace}
\newcommand{\toolname}{\textsc{TestAgent}\xspace}

\usepackage[most]{tcolorbox}

\newcommand{\summary}[2]{
\begin{center}
\begin{tcolorbox}[leftrule=0mm,toprule=0mm,bottomrule=0mm,rightrule=0mm,left=1pt,right=2pt,top=0pt,bottom=0pt,breakable]
\textbf{Answer to RQ{#1}:}
{#2}
\end{tcolorbox}
\end{center}
}

\newcommand{\Description}[1]{}

\AtBeginDocument{
  }

\begin{document}

\title{Multi-Agent LLM Collaboration for Unit Test Generation via Human-Testing-Inspired Workflows}

\author{Quanjun Zhang, Ye Shang, Siqi Gu, Jianyi Zhou, Chunrong Fang,  Zhenyu Chen, Liang Xiao%
\thanks{Quanjun Zhang and Liang Xiao are with Nanjing University of Science and Technology, China. E-mail: quanjunzhang@njust.edu.cn; xiaoliang@mail.njust.edu.cn.}%
\thanks{Ye Shang, Siqi Gu, Chunrong Fang, and Zhenyu Chen are with Nanjing University, China. E-mail: \{yeshang, siqi.gu\}@smail.nju.edu.cn; \{fangchunrong, zychen\}@nju.edu.cn.}%
\thanks{ Jianyi Zhou is with Huawei Cloud Computing Technologies Co., Ltd. E-mail: zhoujianyi2@huawei.com.}
}

\maketitle

\begin{abstract}

Recently, the emergence of Large Language Models (LLMs) has spurred a surge of research into automated unit test generation, yielding impressive performance and reducing manual effort.
However, existing LLM-based approaches still suffer from two major limitations:
(1) they follow rigid, procedural workflows that underutilize the autonomous reasoning potential of LLMs, making it difficult to dynamically adapt testing strategies based on real-time feedback;
and (2) they rely on rule-based context extraction that is not tailored to test generation, failing to capture fine-grained code dependencies and test-specific knowledge required for deriving test requirements.
In this paper, we propose \toolname{}, an LLM-based test generation approach that addresses the above limitations by emulating human testing practices via a multi-agent collaboration mechanism.
Particularly, \toolname{} designs three specialized agents, namely a requirement planner, a test generator, and a test reviewer, to simulate how developers understand, construct, and validate unit tests.
To unleash the autonomous capabilities of LLMs, we equip \toolname{} with a set of tool APIs that can be invoked dynamically in an on-demand and adaptive manner.
To further support repository-level reasoning, \toolname{} constructs a test-specialized knowledge graph via static analysis, which captures code entities and their dependencies across the project and persistently stores testing artifacts (e.g., test reports and failure analyses) produced during generation.
Experimental results show that \toolname{} achieves 97.46\% execution rate, 92.34\% line coverage, 90.24\% branch coverage, and 83.69\% mutation score on six Java projects, outperforming LLM-based baselines across all metrics and achieving substantially higher mutation scores than search-based tools.
We also adapt \toolname{} to Python projects with 88.85\% line coverage and 78.89\% branch coverage, demonstrating its generalizability beyond the Java ecosystem.
Moreover, experiments on industrial projects and a controlled user study confirm the practical applicability of \toolname{} in real-world development scenarios. In addition, \toolname{} detects 154 real-world bugs via non-regression tests with a precision of 92.22\%.
Overall, our study highlights the promising potential of human-testing-inspired multi-agent workflows in producing more reliable, scalable, and practical test cases.

\end{abstract}

\begin{IEEEkeywords}
Software Testing, Test Generation, Large Language Model, AI for SE
\end{IEEEkeywords}

\maketitle

\section{Introduction}
\label{sec:introduction}

Software testing is a cornerstone of modern software quality assurance~\cite{runeson2006survey}.
Among the various stages of testing (e.g., integration and system testing), unit testing plays a particularly critical role by verifying the correctness of individual software components early in the development lifecycle~\cite{olan2003unit, zhu1997software}. 
As the cost of fixing software bugs increases significantly over time, early detection through unit testing substantially reduces maintenance overhead and improves development efficiency.
Thus, unit testing has become a standardized and often mandatory practice as software systems continue to evolve and support a wide range of critical industries~\cite{zhang2024exploring}.

However, it is fundamentally challenging and labor-intensive for developers to construct high-quality unit tests manually~\cite{daka2014survey}.
To mitigate manual efforts in writing unit tests, a variety of techniques have been proposed to automate test generation~\cite{pacheco2007randoop,fraser2011evosuite,fraser2012seed,fraser2013whole}, including heuristic-based~\cite{fraser2011evosuite}, random-based~\cite{pacheco2007randoop}, and symbolic execution~\cite{enoiu2016automated, gargantini1999using, braione2016jbse}.
Despite being promising, test cases generated by these tools typically suffer from limited readability, maintainability, and meaningfulness, which hinders their practical adoption in real-world development.
For example, prior work~\cite{almasi2017industrial} reveals that assertions generated by traditional tools (such as EvoSuite~\cite{fraser2011evosuite} and Randoop~\cite{pacheco2007randoop}) are not as meaningful and helpful as human-written ones in the industrial scenario.

Recently, the rapid advancement of Large Language Models (LLMs) has marked a new milestone in software engineering, demonstrating impressive capabilities in various tasks, such as code generation~\cite{li2023starcoder, wei2024magicoder, wang2021codet5, guo2021graphcodebert, zhang2024codeagent}, summarization~\cite{sun2023prompt, ahmed2024automatic}, and program repair~\cite{zhang2023gamma, zhang2024systematic, zhang2023survey}.
This progress has inspired researchers to explore LLM-based unit test generation from two directions:
(1) training LLMs, which pre-trains or fine-tunes LLMs by collecting high-quality test generation training datasets and selecting efficient training strategies.
(2) prompting LLMs, which constructs effective prompts by extracting code context, retrieving similar test generation examples, and designing task instructions.
Among them, prompt-based techniques, especially those that adopt a generation-and-validation loop, represent state-of-the-art (SOTA) by generating initial test cases and refining them with dynamic feedback.
Despite promising, these techniques still suffer from two major limitations.

\begin{figure*}[t]
    \centering
    \includegraphics[width=0.99\textwidth]{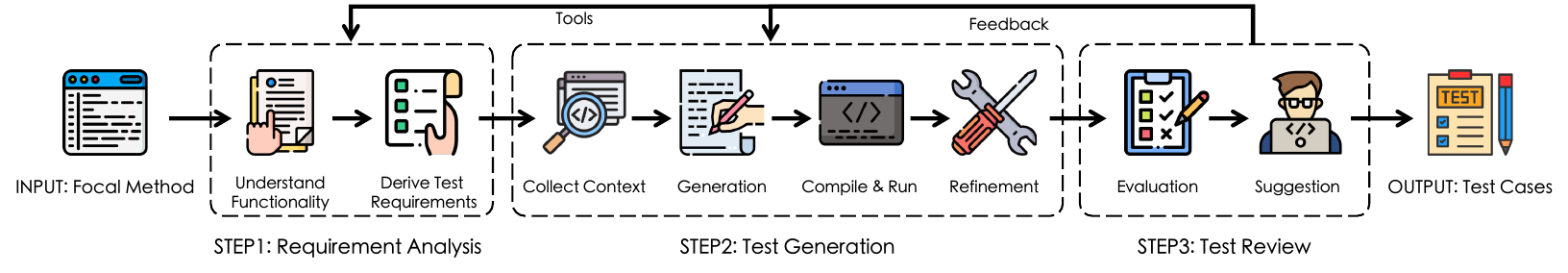}
    \caption{A Common Developer Workflow When Writing Unit Tests}
    \label{fig:developer}
\end{figure*}

\textbf{\ding{182}~Challenge: Procedural Framework.}
Existing approaches typically follow a pre-defined and static generation pipeline that does not adapt the test generation strategy based on real-time feedback (e.g., whether to refine generated test cases or re-extract more relevant context). 
As a result, the generation process remains fixed regardless of intermediate outcomes, limiting its ability to explore alternative generation paths and hampering the effectiveness of LLMs in handling complex or evolving testing requirements.

\textbf{\ding{183}~Challenge: Coarse-grained Context Extraction.}
Most existing approaches retrieve context using fixed rules (e.g., the entire focal method or the focal class), without explicitly modeling language-specific structures such as object-oriented hierarchies, method dispatch, and inter-class dependencies.
For example, extracting only a focal method may omit essential external invocation information, whereas including the entire class or package may introduce irrelevant details and even exceed the input-length limits of LLMs. Although recent SE agents (e.g., RepoGraph~\cite{ouyang2025repograph} and AutoCodeRover~\cite{zhang2024autocoderover}) have explored repository-level code representations to support context retrieval, these representations are designed for general SE tasks such as issue resolution: they neither model test-specific entities and relations (e.g., test classes, test methods, and test--focal links) nor accumulate testing artifacts (e.g., test reports and failure analyses) produced during generation. Consequently, they provide limited support for deriving test requirements and reasoning about the intended behavior of the focal method.

\textbf{This Paper.}
To address these challenges, we propose \toolname{}, a novel LLM-based agent framework for automated unit test generation.
\toolname{} is motivated by the practical experience of developers in writing unit tests, and aims to integrate this domain knowledge with the capabilities of LLMs based on multi-agent collaboration and repository-aware knowledge graphs.
As shown in Figure~\ref{fig:developer}, in real-world unit test generation scenarios, developers typically follow a three-step process: (1) understand the intended program behavior and derive corresponding test requirements; (2) write initial test cases and refine them by execution feedback; and (3) review the quality of test cases with adequacy metrics (e.g., code coverage) and determine whether to deploy them in the production environment.
Taking inspiration from this developer practice, we design \toolname{} with three collaborative agents (i.e., requirement planner, test generator, and test reviewer), each simulating a specific role in the developer's testing workflow.
Particularly, the requirement planner performs semantic analysis of the target function to infer expected behaviors, which is used to generate structured test requirements that specify what aspects of the function should be tested.
The generator agent generates executable test cases with chain-of-thought reasoning and iteratively refines them based on execution feedback in a sandbox environment.
It also conducts root cause analysis on failing test cases and employs a voting-based confirmation mechanism among three different LLMs to determine whether a failure exposes a real bug, providing feedback to guide the next round of test refinement.
The reviewer agent assesses the quality of generated test cases across multiple dimensions (e.g., coverage, correctness, and alignment with testing requirements), and provides actionable suggestions (e.g., deriving new requirements to address uncovered testing scenarios).

To emulate developers' reliance on external tools throughout the testing process, we build a set of tool APIs (e.g., context retrieval and test validation) that agents can invoke autonomously, enabling them to interact with external resources on demand and adjust testing strategies in an adaptive manner, thereby overcoming \textbf{\ding{182}~Challenge} of procedural pipelines.
To overcome \textbf{\ding{183}~Challenge} of context extraction, we construct a knowledge graph via static analysis to model fine-grained dependencies between program entities (e.g., functions, classes, fields, and files) within the whole repository.
Unlike general-purpose repository representations, our knowledge graph is specialized for unit testing: it introduces test-specific nodes and relations (e.g., test classes, test methods, and test--focal links), and further serves as a persistent store for dynamically generated testing artifacts (e.g., function summaries, test reports, and root-cause analyses), allowing agents to accumulate and reuse testing knowledge throughout the generation process.

We conduct extensive experiments to evaluate \toolname{} on three benchmarks, six metrics, five baselines, three underlying LLMs, and two programming languages.
The results show that \toolname{} achieves 97.46\% execution rate, 92.34\% line coverage, 90.24\% branch coverage, and 83.69\% mutation score, consistently outperforming LLM-based baselines (i.e., ChatUniTest and HITS) under the same GPT-4o backbone across all six metrics; compared with the search-based EvoSuite, \toolname{} achieves comparable correctness while delivering substantially higher mutation scores (83.69\% vs. 43.59\%).
The ablation study confirms the impact of core components, e.g., integrating the knowledge graph leads to improvements of 22.31\%, 24.52\%, and 26.83\% in line coverage, branch coverage, and mutation score, respectively.
The experiments with different backbone LLMs demonstrate the strong extensibility of our framework, e.g., \toolname{} with the open-source Qwen3-30B-A3B still outperforming ChatUniTest with GPT-4o, highlighting its strong adaptability without reliance on proprietary models.
We further extend \toolname{} to Python projects, where it achieves 88.85\% line coverage and 78.89\% branch coverage, outperforming dedicated Python-specific LLM-based approaches, including \textsc{CodaMosa} and \textsc{CoverUp}.
We also conduct experiments on industrial projects to demonstrate the effectiveness of \toolname{} in real-world development environments, as well as a user study to demonstrate its practical applicability.
Finally, we evaluate its ability to generate non-regression test cases and find that it successfully detects 154 real-world bugs with 92.22\% precision.

\textbf{Contributions}.
This paper makes the following contributions:
\begin{enumerate}
    \item \textbf{Human-Testing-Inspired Multi-agent Unit Test Generation Framework.}
    We propose \toolname{}, a novel multi-agent framework consisting of a requirement planner, test generator, and test reviewer, which jointly emulate the human testing workflow.
    
    \item \textbf{Structured Code Representation for Agent-based Context Retrieval.}
    We build a repository-level knowledge graph specialized for unit testing, which enables agents to interact with codebases through test-specific nodes and relations (e.g., test classes, test methods, and test--focal links), and persistently stores testing artifacts generated during the process (e.g., function summaries and test reports), providing fine-grained, dependency-aware context throughout test generation.
    
    \item \textbf{Extensive Evaluation.}
    We conduct large-scale experiments with six RQs and two discussions, demonstrating the overall effectiveness of \toolname{}, the impact of its core components, and its extensibility across different underlying LLMs.
    
    \item \textbf{Support for Multiple Languages.}
    We demonstrate the generalizability of \toolname{} across language boundaries by applying it to Python projects, achieving superior performance against LLM-based language-specific approaches.

    \item \textbf{Applicability in Industrial Scenarios.}
    We validate \toolname{} in practical settings through an industrial project evaluation and a user study, confirming its effectiveness and readability.

\end{enumerate}

\section{Background and Related Work}

\subsection{Unit Test Generation}

Existing unit test generation approaches can be broadly categorized into two groups: traditional and LLM-based ones.

\textbf{Traditional Approaches}. 
Traditional approaches typically employ software analysis techniques to automate unit test generation, including heuristic-based~\cite{fraser2011evosuite}, random-based~\cite{pacheco2007randoop}, and symbolic execution~\cite{enoiu2016automated, gargantini1999using, braione2016jbse}.
One prominent approach is search-based software testing (SBST), which uses heuristic evolutionary algorithms, framing test generation as an optimization problem.
Although traditional methods often achieve high code coverage, they tend to generate tests with poor readability and usability~\cite{tufano2020unit, almasi2017industrial}.

\textbf{LLM-based Approaches}. 
Early LLM-based approaches~\cite{tufano2020unit,alagarsamy2024a3test} typically treat the task as a neural machine translation problem with supervised fine-tuning, where the focal method serves as the input, and the output is the generated unit test code.
For example, Tufano et al.~\cite{tufano2020unit} introduce AthenaTest, which utilizes a BART-based model trained in two stages on the Methods2Test dataset~\cite{tufano2022methods2test}. 
Recently, the emergence of powerful closed-source LLMs, such as ChatGPT, has led to increased exploration of prompt engineering approaches for test 
generation~\cite{yuan2024evaluating, chen2024chatunitest, wang2024hits, gu2024testart, pizzorno2024coverup, ni2024casmodatest, lemieux2023codamosa, yang2024enhancing, ryan2024code}.
For example, Chen et al.~\cite{chen2024chatunitest} introduce ChatUniTest, adopting a Generation-Validation-Repair framework to refine generated unit tests iteratively. 
Wang et al.~\cite{wang2024hits} developed HITS, which utilizes LLMs to decompose the focal method into code slices and generate tests incrementally.
Despite promising, current LLM-based approaches still suffer from limitations, such as procedural generation pipelines and coarse-grained context retrieval mechanisms.

\subsection{LLM-based Agents}

The rapid development of LLMs has opened new possibilities for building intelligent agents with broad adaptability. 
Agents, defined as artificial entities capable of perceiving their environment, making decisions, and taking actions, have been considered a promising pathway toward achieving Artificial General Intelligence (AGI)~\cite{bubeck2023sparks}.

The impressive performance of LLM-based agents has motivated extensive research in code-related domains. 
In the code repair domain, Jimenez et al.~\cite{jimenez2024swe} develop SWE-Bench, an evaluation framework comprising 2,294 real-world software engineering issues collected from GitHub. 
It has since become a benchmark for a growing body of agent-based approaches~\cite{lee2024unified, zhang2024autocoderover, yang2024swe, bouzenia2025repairagent, ma2025swe}.
For example, Yang et al.~\cite{yang2024swe} propose SWE-Agent, which employs a customized Agent-Computer Interface to empower agents to create and edit code files, navigate entire repositories, and execute tests.
In the code generation domain, numerous agent-based studies have also been conducted~\cite{arora2024masai, nguyen2024agilecoder, wang2024executable, wang2024rcagent}.

Although sharing the high-level idea of agent-based repository understanding, \toolname{} differs from prior SE agents in three aspects.
First, existing repository representations, such as RepoGraph~\cite{ouyang2025repograph}, are designed for general SE tasks
(e.g., issue resolution) and capture only static code structures; in contrast, our knowledge graph is specialized for unit testing, which introduces test-specific nodes and relations (e.g., test classes, test methods, and test--focal links) and further serves as a persistent store for dynamically generated testing artifacts (e.g., function summaries, test reports, and root-cause analyses).
Second, unlike AutoCodeRover~\cite{zhang2024autocoderover}, which navigates repositories via file paths and rule-based matching, our agent--computer interfaces resolve code entities and traverse inter-procedural relations (e.g., call graphs and inheritance) directly over the knowledge graph, and support dependency-based similar-test retrieval for test reuse.
Third, to the best of our knowledge, \toolname{} is the first work to incorporate requirement planning, test generation, and test reviewing into a multi-agent unit testing pipeline that emulates the human testing workflow.

\section{Methodology}
\label{sec:methodology}

\begin{figure*}[t]
  \centering
  \includegraphics[width=0.95\linewidth]{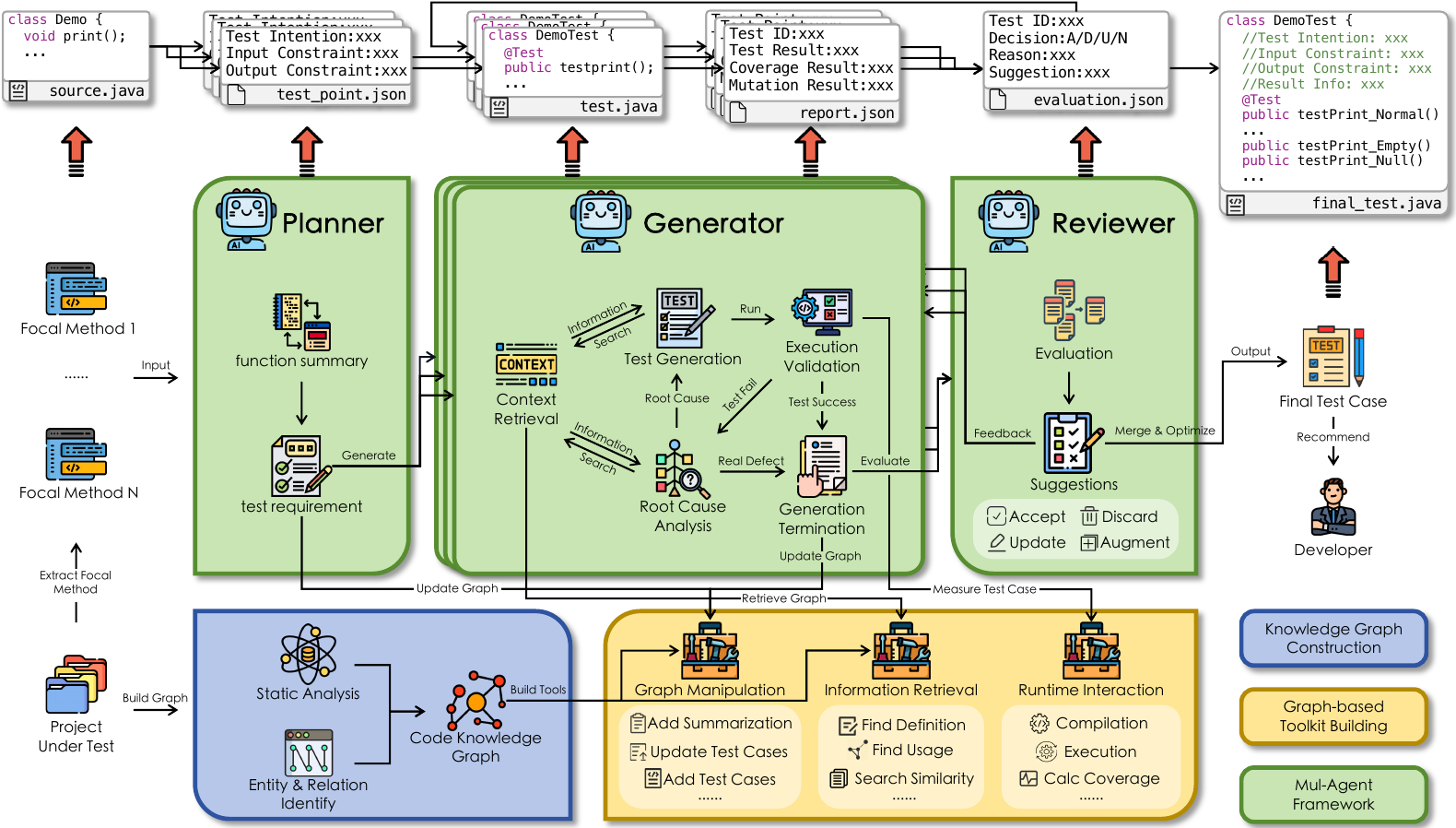}
  \caption{Overview of \toolname{}}
  \label{fig:workflow}
\end{figure*}

\subsection{Overview}

Figure~\ref{fig:workflow} presents the overall workflow of \toolname{}, an LLM-agent-based framework for unit test generation.
Given a focal method under test and its project repository, \toolname{} consists of three primary components: 
(1) a knowledge graph for repository-level code understanding, (2) a set of tools that orchestrate the communication between agents and the project environment, and (3) a multi-agent collaboration framework that interactively synthesizes, refines, and reviews high-quality unit tests.

First, \toolname{} constructs a knowledge graph by utilizing static program analysis to extract program entities and semantic relationships.
The structured graph represents class hierarchies, method invocations, and variable dependencies, forming the semantic foundation for repository understanding and agent reasoning.
Second, \toolname{} builds a set of agent-computer interfaces based on the knowledge graph, enabling agents to efficiently query repository-level context information. 
These tools provide dynamic and fine-grained access to the whole project repository, supporting precise context acquisition during the test requirement modeling, generation, and evaluation stages.
Third, \toolname{} designs a three-stage multi-agent framework with each stage coordinated by a dedicated agent.
The requirement planner performs semantic analysis of the method under test and derives corresponding test requirements that specify what aspects of the functionality should be tested.
The test generator produces syntactically correct and executable test cases from the code context and derived requirements.
It also performs failure root cause analysis and employs a voting-based confirmation mechanism to identify whether the failure detects actual defects in the target function.
The test reviewer assesses test cases across multiple dimensions and provides actionable suggestions to guide the subsequent behavior of other agents, enabling continuous iteration.

\subsection{Knowledge Graph Construction}

To support repository-level reasoning for LLMs, we construct an entity-level knowledge graph based on static code analysis to capture the internal semantics and relationships within the project.
This graph serves as an explicit, queryable representation that bridges the gap between raw source code and LLMs.
The input to this process is the complete code repository, including all its folders and files, while the output is a structured graph, whose nodes correspond to program entities (e.g., packages, classes, methods, variables), and edges encode semantic dependencies among these entities. 
Formally, given a project $\mathcal{P}$ composed of source files $p_i \in \mathcal{P}$, the construction process involves three main steps:

\textbf{\ding{172} Code Entity Extraction.}
For each file $p_i$, we generate its abstract syntax tree (AST) using SPOON~\cite{pawlak2016spoon}.
{
\begin{equation}
    AST_i = \text{BuildAST}(p_i)
\end{equation}
}
Then, we hierarchically extract the set of entities $\mathcal{E}$ by traversing all ASTs.
{
\begin{equation}
    \mathcal{E} = \bigcup_{i} \text{ExtractEntities}(AST_i)
\end{equation}
}
These entities are organized into four primary categories, ranging from coarse-grained modules to fine-grained elements: packages, classes, methods, and variables.
Each category is further refined based on its programming semantics.
For example, class entities are classified into subtypes, such as regular classes, abstract classes, interfaces, enumerations, decorators, and test classes. 

\textbf{\ding{173} Dependency Relation Extraction.}
After recognizing entity definitions, we identify inter-entity dependencies $\mathcal{R}$ within the project via Intermediate Representation (IR) and pointer analysis.
{
\begin{equation}
    \mathcal{R} = \bigcup_{e \in \mathcal{E}} \text{AnalyzeDependencies}(\text{IR}(e), \mathcal{E})
\end{equation}}

To capture semantic relationships among these entities, we define a rich set of dependency types in $\mathcal{R}$, including inheritance, implementation, method invocation, and reference, reflecting core semantic structures in real-world software.

\textbf{\ding{174} Graph Assembly.}
Finally, we construct the knowledge graph $\mathcal{KG}$ as a set of $(h, r, t)$ triples, where $h$ and $t$ are entities and $r$ denotes their relationship:
{
\begin{equation}
    \mathcal{KG} = \{(h, r, t) \mid h, t \in \mathcal{E}, r \in \mathcal{R}\}
\end{equation}
}

The graph is constructed once per project and shared across all focal methods, and is incrementally maintained by the graph manipulation tools as testing proceeds.

\subsection{Graph-based Toolkit Building}

\input{tables/tools}

Building on the knowledge graph that serves as a structured representation of the whole repository, we design a set of tools to facilitate efficient interaction between agents and external resources.

This toolkit provides agents with accurate and efficient access to the entire repository during the testing process.
Overall, it supports three types of tools:
\textbf{(1) Graph Manipulation Tools.}
These tools enable agents to update the knowledge graph based on testing progress. 
For example, the requirement planner agent can invoke \texttt{add\_method\_summarization} to insert semantic summaries for a specific focal method.
\textbf{(2) Information Retrieval Tools.}
These tools enable agents to extract fine-grained repository-level context from the knowledge graph.
For example, entity lookup tools (e.g., \texttt{find\_method\_definition} and \texttt{find\_variable\_definition}) support precise queries based on fully qualified names and function signatures to ensure disambiguation.
Dependency Tracing Tools (e.g., \texttt{find\_method\_calls}) return caller and callee relationships for a given function entity.
Fuzzy matching tools (e.g., \texttt{fuzzy\_search}) support approximate queries when exact identifiers are unavailable, returning top-ranked candidates based on semantic similarity.
Test Recommendation tools (e.g., \texttt{search\_similar\_test\_class}) identify and rank semantically similar test cases for reuse or adaptation.
\textbf{(3) Runtime Interaction Tools.}
These tools enable agents to validate the generated test cases via dynamic analysis, including correctness checks (\texttt{check\_syntax} and \texttt{compile\_test\_cases}) and adequacy checks (e.g., \texttt{calculate\_coverage}).

Due to space limitations, we omit the detailed implementation of all individual tools.
Instead, we illustrate one representative tool, \texttt{search\_similar\_test\_class}, in detail, which aims to recommend related test cases based on an entity dependency-based similarity metric defined over the knowledge graph. 
The intuition is that tests sharing dependencies tend to share testing logic: tests targeting different implementations of one interface often follow the same testing logic despite divergent surface code, whereas lexically similar tests may exercise unrelated behavior. 
We therefore measure similarity over the entities a test depends on, rather than over its surface code.
Given a target test case entity $tc$ and a knowledge graph $\mathcal{KG}$, we first extract all other test case entities as the candidate set: $\mathcal{TC} = \{e \in \mathcal{E} \mid \text{type}(e) = \text{TestCase}, e \ne tc\}$,
where $\mathcal{E}$ denotes all entities in $\mathcal{KG}$, and $\text{type}(e)$ returns the type of entity $e$.
Then, we compute the dependency set for $tc$ based on its outgoing relations:
{
\begin{equation}
    D(tc) = \{(r, e') \mid (tc, r, e') \in \mathcal{KG}\}
\end{equation}}
where $r \in \mathcal{R}$ is the type of relation (e.g., calls, defines, uses), and $e' \in \mathcal{E}$ is the related entity.
Similarly, we extract dependency sets for each candidate test case $t \in \mathcal{TC}$: $D(t) = \{(r, e') \mid (t, r, e') \in \mathcal{KG}\}$.
We then compute the Jaccard similarity between $tc$ and each $t$: $\text{sim}(tc, t) = \frac{|D(tc) \cap D(t)|}{|D(tc) \cup D(t)|}$.
Finally, all candidate test cases are ranked by similarity, and those with $\text{sim}(tc, t) \geq k$ (for a predefined threshold $k$) are selected as the recommendation set:

{
\begin{equation}
    \mathcal{TC}_{\text{related}} = \{t \mid t \in \mathcal{TC}, \text{sim}(tc, t) \geq k \}
\end{equation}
}

All three types of tools are exposed to every agent, which autonomously selects the appropriate tool according to its current state.

\subsection{Multi-Agent Framework}
To autonomously guide LLMs to leverage knowledge graphs and external tools, motivated by the three-stage unit test generation process typically employed by developers, we design three dedicated agents (i.e., requirement planner, test generator, and reviewer) to simulate this development practice. 
Each agent is responsible for a specific stage of the process, reflecting the natural division of tasks in human-driven test creation.

\subsubsection{Requirement Planner Agent}
The requirement planner agent is responsible for analyzing the semantic behavior of the target function and planning structured test requirements that guide LLMs to explore meaningful and diverse test scenarios.
To automate this process, this agent operates in two internal states: \textbf{function summary} and \textbf{test requirement}.

In the function summary state, \toolname{} designs a prompt template to conduct semantic analysis and abstraction of the target function
The output is a structured function summary that captures the function’s purpose, input parameters, return values, and notable behaviors.
Once the summary is generated, \toolname{} invokes the knowledge graph update interface to associate the summary with the corresponding function entity, ensuring that semantic information is readily retrievable in later stages.

In the test requirement analysis state, given the function summary as input, \toolname{} analyzes the function across three testing dimensions: normal inputs, boundary cases, and exceptional scenarios. 
It considers semantic roles, parameter types, edge behaviors, and exception handling to construct a comprehensive set of test requirement points.
Each test requirement point specifies the testing intention, input type, and expected output behavior.

\subsubsection{Test Generator Agent}

The test generator agent is responsible for producing and iteratively refining high-quality unit test cases that fulfill the specified test requirements.
To automate this process, the agent operates through a series of adaptive states based on the ``observe-think-act'' cognitive decision-making paradigm.
Given a focal function and its test requirements, the agent progresses through the following major states:

\ding{182} \textbf{Context Retrieval}.
The agent begins by utilizing an external retrieval toolkit to actively query context information (e.g., program entities and dependencies) related to the focal method from the knowledge graph.
The agent then assesses whether the current context is sufficient to support meaningful test generation. 
If so, it transitions to the next state; otherwise, it remains in the retrieval state to gather additional context information.
To bound this loop, \toolname{} adopts an elastic control strategy: the agent is explicitly queried about context sufficiency after 30 interaction turns, and test generation is enforced after 40 turns, within the maximum recursion depth described in Section~\ref{sec:experiment_setup}.

\ding{183} \textbf{Test Generation}.
Upon confirming adequate context, the agent synthesizes candidate test cases using a chain-of-thought-based reasoning prompt.
This prompt guides the agent through step-by-step reasoning to construct structurally sound and semantically meaningful test cases that fulfill the given test requirements.

\ding{184} \textbf{Execution Validation}.
The correctness of generated test cases is verified via an automated pipeline, in which the agent invokes external tools for syntax checking, compilation, and sandboxed execution. 
Test cases that pass all checks are directly added to the final suite; those that fail trigger the agent to enter the root cause analysis stage.

\ding{185} \textbf{Root Cause Analysis}.
For any failing test case, the agent performs root cause analysis to determine the underlying reason:
(1) Test Case Issue.
If the failure is caused by minor flaws in the test case, such as incorrect arguments, weak assertions, or improper handling of edge conditions, the agent considers it recoverable.
In this case, the test case is refined and re-entered into the generation-validation loop with detailed validation feedback.
(2) Focal Method Issue.
If the failure is likely caused by a defect in the function under test, the agent activates a defect confirmation mechanism.
The mechanism invokes three LLMs with different training pipelines, each of which independently examines the test case, error message, and reasoning trace. 
If all three agree that the failure reveals a real defect, the test case is labeled as defect-revealing and added to the final test suite.

\ding{186} \textbf{Generation Termination}.
Once a test case either successfully passes all validations or is confirmed to expose a real defect, the agent generates a structured report to record the testing outcomes, including line coverage, branch coverage, and mutation scores. 
These results are formatted into a JSON format report, which serves as feedback for the reviewer agent or as documentation for developer inspection when deployment.
If the retry budget is exhausted before a compilable test is produced for a highly complex focal method, the agent records the last attempt as a best-effort result rather than discarding the method.

\subsubsection{Test Reviewer Agent}
The test reviewer agent is responsible for assessing the quality of generated test cases from multiple perspectives, and providing actionable suggestions that are iteratively fed back to the above two agents to guide their subsequent behaviors of both test cases and test requirements.
To automate this process, this agent operates in two internal states: {quality evaluation} and {suggestion derivation}.

In the quality evaluation state, the agent receives the full context required for assessment, including the function definition, function summary, related test requirements, generated test cases, and execution reports.
It then analyzes each test case based on multiple criteria, including syntax validity, compilation success, runtime behavior, structural coverage (e.g., line and branch coverage), alignment with test requirements, and defect detection capabilities (e.g., mutation score), to generate an evaluation summary.

In the feedback analysis state, the agent translates evaluation results into actionable decisions for both test cases and test requirements. 
For each test case, it assigns one of four feedback actions:
(1) Accept. The test case satisfies the expected requirement, achieves meaningful coverage, and exhibits strong fault detection (e.g., kills mutants or reveals defects).
Thus, it is retained for deployment without modification.
(2) Discard. The test case fails to meet the intended requirement and contains unrecoverable flaws.
Thus, it is removed from the generated test suite, and its related graph entries are deleted (i.e., calling \texttt{delete\_test\_cases} tool in Table~\ref{tab:tools}).
(3) Update. The test case is partially valuable but incomplete, e.g., minor syntax or compilation errors, partial coverage, or weak assertions. 
Thus, it is retained with concrete suggestions for improvement.
(4) Augment. If a test case exhibits inadequate coverage or fault detection due to insufficient test requirements, the agent infers new test requirements targeting untested scenarios, such as uncovered paths, unhandled exceptions, or surviving mutants.

\subsection{Implementation}
We implement \toolname{} as a VSCode unit generation plugin to support practical adoption and developer interaction in real-world workflows.
Developers can initiate the plugin by selecting a focal method or an entire project within the editor.
The plugin outputs (1) executable test cases with high coverage and mutation score, suitable for direct use as regression tests, and (2) bug-revealing failing tests for inspecting potential defects in current project versions.
A JSON report is also provided, detailing each test case, its target requirement, coverage, and mutation performance, facilitating transparent inspection and CI integration.
All intermediate agent reasoning steps are displayed in an interactive panel, where developers can preview, edit, or insert test cases as needed.

\section{Experiments and Results}
\label{sec:results}

\subsection{Experiment Setup}
\label{sec:experiment_setup}

\subsubsection{Research Questions}
Our evaluation investigates the following research questions (RQs).

\textbf{RQ1: Effectiveness Comparison.}
    How does \toolname{} perform compared to baselines on standard metrics, including correctness, coverage, and mutation?

\textbf{RQ2: Ablation Study.} 
    How do individual components contribute to the overall performance of \toolname{}, including test retrieval, test reviewer, and knowledge graph?

\textbf{RQ3: Model Adaptability.} 
    How does \toolname{} perform across different underlying LLMs?

\textbf{RQ4: Industrial Project Performance.} 
    How effective is \toolname{} in generating unit tests for real-world industrial projects?

\textbf{RQ5: Readability and Usability.} 
    Do test cases from \toolname{} offer higher practicality than those from baselines?

\textbf{RQ6: Language Generalizability.} 
    Can \toolname{} maintain its effectiveness in generating test cases for Python?

\input{tables/dataset}

\subsubsection{Dataset}
We evaluate \toolname{} using three distinct datasets to comprehensively assess its performance from multiple perspectives.
First, in RQ1-RQ3, consistent with prior studies, we collect six Java projects from GitHub, resulting in a dataset containing 1,454 focal methods. 
Five of these projects originate from Defects4J~\cite{just2014defects4j}, a widely recognized benchmark dataset frequently utilized in previous evaluations, and we adopt the same five projects used in prior LLM-based test generation studies~\cite{alagarsamy2024a3test, chen2024chatunitest, wang2024hits} to ensure a consistent comparison with baselines. Due to resource constraints, we randomly select 200 methods each from the Lang and Chart projects, both of which originally contain thousands of methods. 
To evaluate the performance of \toolname{} in complex methods, we also include another project, Ruler, identified by HITS as particularly challenging.
Second, in RQ4,  we introduce an internal dataset named UTXXX, to assess how \toolname{} performs on previously unseen industrial methods.
This dataset, sourced from industry, comprises 658 focal methods collected from four Java projects.
Finally, in RQ6, to evaluate the applicability of \toolname{} to Python programs, we follow the approach of \textsc{CodaMosa} and select nine modules from three Python projects. These modules form a Python dataset, referred to as CM, containing a total of 141 focal methods.
Since the open-source subjects are mainly library projects, the industrial dataset in RQ4 further complements them with projects that better reflect industrial practice.
The details and statistics of these datasets are summarized in Table~\ref{tab:dataset-info}.

\subsubsection{Baselines}

To address RQ1-RQ5, we compare \toolname{} against three state-of-the-art baselines:
\textbf{(1) EvoSuite}~\cite{fraser2011evosuite} combines static analysis with genetic algorithms to maximize coverage, serving as a strong baseline in both academic and industry evaluations.
\textbf{(2) ChatUniTest}~\cite{chen2024chatunitest} incorporates adaptive focal context to provide semantic grounding, and utilizes a generate–validate–repair loop to ensure test correctness and executability.
\textbf{(3) HITS}~\cite{wang2024hits} decomposes each complex focal method into smaller program slices and prompts the LLM to generate tests for every slice independently.

To answer RQ6, we implement \toolname{} in Python and compare it with two LLM-based Python-specific baselines:
\textbf{(1) \textsc{CodaMosa}}~\cite{lemieux2023codamosa} enhances SBST by querying LLMs for seed tests when coverage plateaus, guiding exploration toward uncovered paths.
\textbf{(2) \textsc{CoverUp}}~\cite{pizzorno2024coverup} iteratively prompts LLMs with execution feedback and uncovered code regions until coverage convergence.

\subsubsection{Metric}
\label{sec:metric}
To evaluate the effectiveness of the generated tests, we employ six quantitative metrics:
\textbf{(1) Syntax correctness rate (Sy. Rate)}: The proportion of generated tests with correct syntax.
\textbf{(2) Compilation success rate (Co. Rate)}: The proportion of generated tests that compile without errors.
\textbf{(3) Execution success rate (Exe. Rate)}: The proportion of tests that run successfully without runtime errors.
\textbf{(4) Line coverage (Li Cov.)}: the proportion of code lines executed.
\textbf{(5) Branch coverage (Br Cov.)}: the proportion of code branches executed.
\textbf{(6) Mutation score (Mut Sc.)}: The effectiveness of tests in detecting artificially injected faults.

To investigate developer-oriented feedback, following \textsc{ChatTester}, we introduce four \textbf{Readability and Usability} metrics in RQ5: naming intuitiveness (NI), code layout (CL), assertion quality (AQ), and adoption effort (AE)—aggregated.
To explore the capability of detecting real-world defects, we conduct a \textbf{Bug-Detection Effectiveness} analysis in the discussion section. This is measured by precision ($P = \frac{TP}{TP + FP}$), with confirmed true positives further categorized along two orthogonal axes: root cause and observed impact, each divided into five subtypes (see Table~\ref{tab:root_analysis}).
We also consider two metrics: \textbf{Monetary and Time Costs} to evaluate the cost-effectiveness of \toolname{}.

\subsubsection{Configuration Setting}
We primarily evaluate \toolname{} using OpenAI's GPT-4o~\cite{hurst2024gpt} (\texttt{gpt-4o-2024-05-13}) as the base LLM.
For a fair comparison, the LLM-based baselines (i.e., ChatUniTest and HITS) are executed with the same GPT-4o version as \toolname{}, and all baselines adopt the configurations recommended by their original authors.
In particular, EvoSuite is run with a search budget of 300 seconds per class. 
To assess the model-agnostic capability, we evaluate \toolname{} using two alternative LLMs: DeepSeek-V3~\cite{deepseekai2024deepseek} and Qwen3-30B-A3B~\cite{yang2025qwen3}. 
All models utilize default parameters without limiting output tokens. Specifically, Qwen3-30B-A3B is deployed using vLLM on two Nvidia A100 GPUs.
In experiments, we employ Neo4J~\cite{webber2012programmatic} for graph management.
To prevent long contexts, we select only the top-5 test requirements generated by the planner agent for subsequent processing, and set the maximum recursion depth to 50.
The defect confirmation mechanism employs o3-mini, DeepSeek-R1, and
Qwen3-235B-A22B as the three voting models.

\subsection{RQ1: Effectiveness Comparison}

\input{tables/rq1_performance}

\textbf{Design}. 
RQ1 aims to investigate the overall effectiveness of the test cases generated by the \toolname{}.
We consider six metrics and compare against three SOTA baselines (EvoSuite, ChatUniTest, and HITS) to comprehensively evaluate the performance of \toolname{}.

\textbf{Results and Analysis}.
Table~\ref{tab:overall_performance} presents the comparison results of \toolname{} and three baselines across six Java projects.
Overall, \toolname{} demonstrates superior performance compared to the baseline approaches. 
Although EvoSuite achieves a perfect compilation rate (100.00\%) and a high execution rate (98.07\%), \toolname{} closely matches these results with a compilation rate of 99.11\% and an execution rate of 97.46\%, substantially surpassing ChatUniTest (81.29\% compilation, 66.09\% execution) and HITS (91.54\% compilation, 81.71\% execution).
The residual compilation failures (0.89\%) arise from highly complex focal methods whose retry budget is exhausted, for which \toolname{} records the last attempt as a best-effort result.
Moreover, \toolname{} attains substantially higher line coverage (92.34\%), branch coverage (90.24\%), and mutation score (83.69\%) compared to EvoSuite (82.11\%, 80.81\%, and 43.59\%), ChatUniTest (72.40\%, 71.44\%, and 51.79\%), and HITS (82.39\%, 81.22\%, and 63.14\%), highlighting its effectiveness in generating high-quality test cases that effectively detect faults.

Regarding individual projects, we observe that some projects (e.g., Cli, Csv, Chart, and Lang) are relatively straightforward, while others (e.g., Gson and Ruler) present greater challenges. Nevertheless, even on these more challenging projects, \toolname{} consistently achieves robust performance, demonstrating its adaptability and effectiveness across diverse project complexities.

\summary{1}{ 
\toolname{} consistently outperforms LLM-based baselines across all six metrics, and achieves comparable correctness with substantially higher coverage and mutation scores than the search-based EvoSuite.
Its consistent performance across diverse projects, including challenging scenarios, highlights its robustness and practical applicability.
}

\subsection{RQ2: Ablation Study}

\textbf{Design}.
RQ2 explores the individual contribution of each major component within \toolname{}.
We conduct an ablation study by removing (i) similar test retrieval, (ii) test reviewer, and (iii) knowledge graph, which correspond to three architectural levels of \toolname{}: a representative retrieval tool, a collaborative agent, and the underlying repository representation, yielding three variants of the system.

\input{tables/rq2_ablation}

\textbf{Results and Analysis}.
Table~\ref{tab:ablation-study} reports the results for the full \toolname{} and its three ablated variants. 
Syntax correctness remains 100\% under all configurations, confirming that the generator consistently produces passing test cases. 
Nevertheless, omitting specific components degrades the remaining metrics to varying degrees.
First, removing the retrieval module causes modest declines, e.g., compilation success drops by 0.50\%, execution success by 0.64\%, line and branch coverage by roughly 1–2\%, and mutation score by 2.87\%. 
Second, eliminating the reviewer produces a larger deficit, e.g., execution success falls by 2.69\%, line and branch coverage decline by 4.52\% and 5.25\%, respectively, and mutation score drops by 8.09\%. 
These results show that the feedback loop implemented by the reviewer is essential for filtering ineffective tests and steering generations toward the intended behaviour.
Third, the most severe degradation arises when the knowledge-graph module is removed. 
Execution success plunges by 12.11\%, line and branch coverage fall to 75.65\% and 72.62\%, and the mutation score sinks to 66.12\%. These findings underscore the central role of the knowledge graph in modelling program context and dependencies, which substantially broadens coverage and strengthens fault detection in multi-module or dependency-rich code bases.

\summary{2}{
All components contribute positively to the overall effectiveness of \toolname{}.
Notably, the knowledge graph drives the largest gains, as its structured view of inter-procedural dependencies provides LLMs with precise context.
}

\subsection{RQ3: Model Adaptability of \toolname{}}
\textbf{Design.}
RQ3 investigates the generalizability of \toolname{} across different base LLMs. We select three representative LLMs: GPT-4o, DeepSeek-V3, and Qwen3-30B-A3B. These models include both closed-source and open-source LLMs with varying model sizes, accessed via API or deployed on independent hardware. This setup demonstrates the model-agnostic characteristic of \toolname{} and its adaptability to diverse deployment scenarios.

\input{tables/rq3_models}

\textbf{Results and Analysis}.
Table~\ref{tab:model_compare} summarizes the performance comparison among the selected LLMs. GPT-4o consistently achieves the best overall performance, followed by DeepSeek-V3, with Qwen3-30B-A3B showing the lowest performance. This ranking aligns with their corresponding deployment costs and capabilities.
Regarding test correctness, all three models achieve a syntax correctness rate of 100\%, indicating robust language modeling abilities. Regarding compilation and execution success rates, GPT-4o and DeepSeek-V3 both exceed 97\%, whereas Qwen3-30B-A3B shows a noticeable performance degradation, with compilation success dropping to 94.57\% and execution success declining further to 85.63\%.
Regarding test adequacy, GPT-4o leads substantially, achieving 92.34\% line coverage, 90.24\% branch coverage, and 83.69\% mutation score, reflecting superior capability in generating high-quality test cases. 
DeepSeek-V3 trails GPT-4o by approximately 6\% in both line and branch coverage, and by roughly 11\% in mutation score. 
Qwen3-30B-A3B exhibits the lowest adequacy, with 81.25\% line coverage, 78.21\% branch coverage, and 60.44\% mutation score.
However, it still substantially surpasses ChatUniTest (even when using GPT-4o), highlighting the effectiveness of \toolname{} regardless of the underlying LLM.

We also observe that \toolname{} with Qwen3-30B-A3B trails EvoSuite on line and branch coverage (by 0.86\% and 2.60\%, respectively), which stems mainly from the limited context understanding and requirement analysis capability of this smaller open-source model.
Nevertheless, it remains well ahead of EvoSuite on mutation score (60.44\% vs. 43.59\%), while incurring zero API cost due to local deployment.
In this light, we regard \toolname{} and EvoSuite as complementary rather than strictly competing approaches: EvoSuite excels at maximizing structural coverage at very low cost, whereas \toolname{} contributes higher mutation scores and more readable, requirement-aligned tests (see RQ5).
Combining the two, e.g., seeding \toolname{} with EvoSuite-generated tests to close residual coverage gaps, is a promising direction for future work.

\summary{3}{
\toolname{} exhibits strong adaptability across diverse LLM backends. 
While performance correlates with model strength, all evaluated models benefit from \toolname{}'s framework. This confirms \toolname{}’s model-agnostic design and its robustness in varied deployment scenarios.
}

\subsection{RQ4: Industrial Project Performance}
\textbf{Design}.
To mitigate potential data leakage risk, RQ4 assesses the performance of \toolname{} and baseline methods using UTXXX, a real-world industrial dataset comprising 658 target methods from four distinct projects. 
Due to the confidential policy of the company, we hide the information about internal subjects. The UTXXX adopts a unified Maven architecture with the JUnit 4 testing framework. The projects have never been publicly released, substantially reducing the risk of exposure during LLM pre-training. This setup aligns with mainstream development practices and provides a more accurate measure of capability for generating test cases in contemporary Java projects. 
Specifically, we use the Qwen3-30B-A3B hosted on the local server. However, due to limitations in this model's capability, it could not perform code slicing—a crucial step required by the HITS approach. Consequently, we excluded HITS from this comparison. To objectively measure the effectiveness of each testing approach, we used three widely adopted metrics: line coverage, branch coverage, and mutation score.

\input{tables/disc_industry}

\textbf{Results and Analysis}.
Table~\ref{tab:industry_benchmark} presents the performance results of each approach evaluated on the UTXXX dataset. As shown, \toolname{} consistently achieves the highest performance across all three metrics. Specifically, \toolname{} attains a line coverage of 84.21\%, branch coverage of 82.12\%, and mutation score of 55.38\%, clearly outperforming the other methods. 
ChatUniTest exhibits comparable performance to EvoSuite in terms of line coverage and branch coverage, but achieves a noticeably higher mutation score. 
Overall, these results demonstrate that \toolname{} substantially outperforms baseline approaches, highlighting its effectiveness in generating high-quality unit test cases for industrial Java projects.

\summary{4}{
\toolname{} demonstrates strong practical effectiveness in real-world industrial projects. 
Despite being equipped with a smaller LLM (Qwen3-30B-A3B), it substantially outperforms established baselines in coverage and mutation scores, showcasing its robustness and real-world applicability.
}

\subsection{RQ5: Readability and Usability Evaluation}

\textbf{Design}.
RQ5 evaluates the readability and usability of test cases generated by \toolname{} and baselines (i.e., EvoSuite, ChatUniTest, and HITS) through a user study, following the methodology of \textsc{ChatTester}~\cite{yuan2024evaluating}. 
Specifically, we randomly select 100 focal methods, with executable test cases generated by all four approaches, resulting in a total of 400 test cases. 
We invite ten participants, each with 3-5 years of Java development experience, to independently assess test cases based on four criteria, i.e., naming intuitiveness and code layout for readability, and assertion quality and adoption effort for usability.
Each criterion is scored on a 1–3 scale, with higher scores indicating better quality. 
To ensure impartiality, participants are blinded to the source of each test case.

\input{tables/disc_readable}

\textbf{Results and Analysis}.
Table~\ref{tab:readability_results} presents the comparison results of \toolname{} and baselines.
Overall, \toolname{} achieves the highest readability and usability scores. 
ChatUniTest ranks second, obtaining the highest code layout score primarily due to its concise test cases. 
In contrast, HITS receives lower scores, mainly due to unnecessary import statements and unclear code structures. EvoSuite performs the poorest, mainly due to insufficient comments, meaningless string values, and unclear testing intention.

\summary{5}{
Our user study shows that test cases generated by \toolname{} consistently achieve the highest readability and usability scores, indicating that \toolname{} produces not only effective but also developer-friendly tests.
}

\subsection{RQ6: Language Generalizability}

\textbf{Design}.
In this work, we implement \toolname{} in Java as it is the most widely adopted language in unit testing, offering a rich ecosystem of benchmark projects and strong baselines for comparison.
In fact, the core methodology of \toolname{} is language-agnostic and can be extended to other programming languages with minimal engineering effort.
The primary adaptation lies in the knowledge graph construction phase, where a language-specific AST parser is required to extract code entities and relationships.
To facilitate a fair comparison with SOTA Python-based test generation approaches such as \textsc{CodaMosa} and \textsc{CoverUp}, we implement a Python version of \toolname{}. Considering the differences between Python and Java, we redesign the knowledge graph for Python, remove components related to compilation, and adapt other modules accordingly to ensure compatibility with Python.

\input{tables/disc_python_result}

\textbf{Results and Analysis}.
We evaluate all methods on nine modules from three Python projects, where both \textsc{CodaMosa} and \textsc{CoverUp} perform suboptimally. The results are presented in Table~\ref{tab:coverage-comparison}. 
\toolname{} consistently achieves the highest performance across all nine modules. On average, it improves line coverage by 26.86\% and branch coverage by 50.81\% compared to \textsc{CoverUp}, the strongest baseline. Although our primary experiments focus on Java projects, this supplementary study demonstrates that \toolname{} remains effective in Python environments as well. These findings suggest that the core principles underlying \toolname{} generalize well across programming languages, indicating its potential as a language-agnostic test generation framework.

\summary{6}{
\toolname{} exhibits strong cross-language adaptability. When reimplemented for Python, it substantially outperforms SOTA Python baselines like \textsc{CodaMosa} and \textsc{CoverUp} in both line and branch coverage. This confirms its potential as a general-purpose, language-agnostic test generation framework.
}

\section{Discussion}

\subsection{Cost-Effectiveness Analysis}

\input{tables/disc_bug_detect_desc}

\begin{figure}[t]
  \centering
  \includegraphics[width=0.99\linewidth]{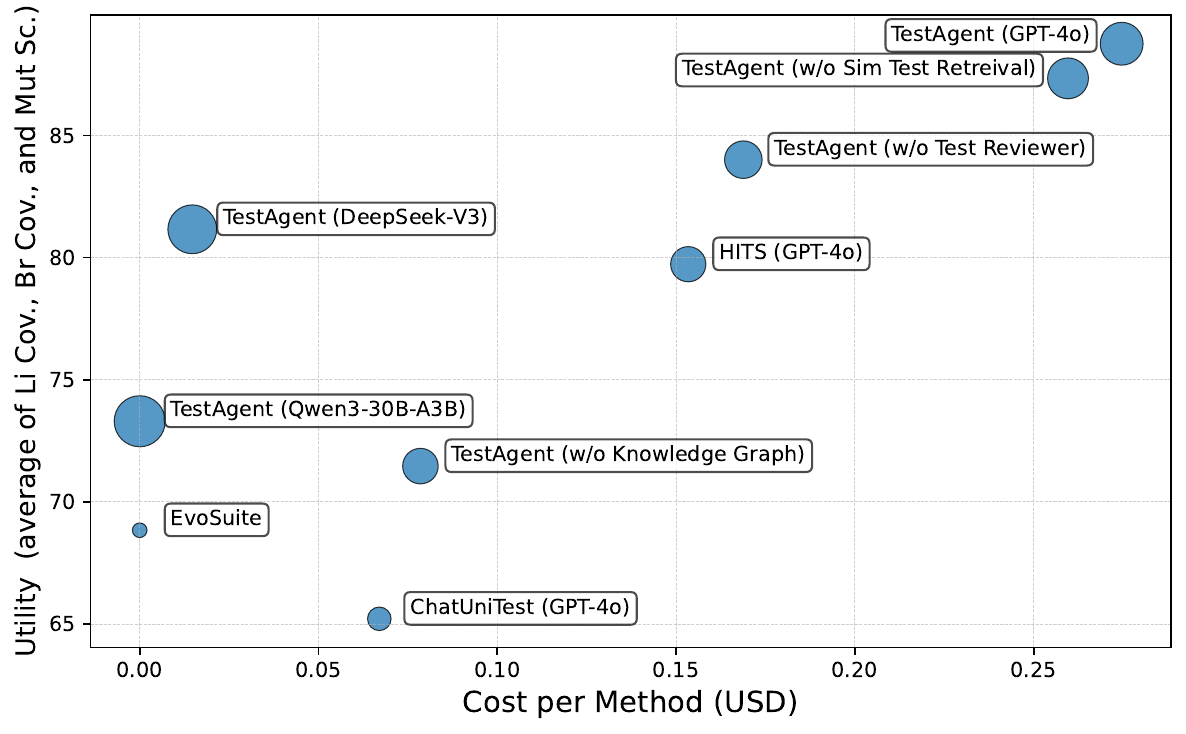}
  \caption{Cost-Effectiveness of \toolname{} Variants}
  \label{fig:rq4_scatter}
\end{figure}

\textbf{Design}.
Building upon the findings from RQ2 and RQ3, we further analyze the cost-effectiveness of various configurations of \toolname{}. Specifically, we analyze monetary and time costs across different variants, quantifying their performance (utility) as the average of line coverage, branch coverage, and mutation score.

\textbf{Results and Analysis}.
Figure~\ref{fig:rq4_scatter} illustrates the cost-effectiveness landscape of different test-generation approaches, and the larger size of the scatter means the longer time costs.
The results show that \toolname{} with GPT-4o achieves the highest utility, but incurs the greatest cost per method. Variants with reduced components, such as the omission of similar test retrieval or evaluation phases, lower the cost considerably but sacrifice some performance. 
Among these simplified variants, \toolname{} without the test reviewer achieves a better balance between cost reduction and performance retention.
Notably, \toolname{} with DeepSeek-V3 demonstrates exceptional cost-effectiveness, maintaining 91.44\% of the performance of GPT-4o while reducing monetary cost to merely 5.35\%. 
Moreover, \toolname{} with Qwen3, characterized by zero API monetary cost due to local deployment, still outperforms EvoSuite substantially, although it requires more computational time (approximately 300 seconds per method compared to EvoSuite's 19 seconds per method). The variants without the knowledge graph and ChatUniTest show the poorest utility among the evaluated configurations.

\subsection{Bug Detection via Non-regression Tests}
\textbf{Design}.
As discussed in Section~\ref{sec:results}, we have demonstrated that \toolname{} achieves high mutation scores, indicating strong potential in generating regression tests that expose bugs by comparing behavior across versions.
In this section, we evaluate \toolname{}'s ability to generate non-regression tests, i.e., tests that detect faults directly in the current version of the code.
Unlike existing approaches that treat all failures as test code issues and repeatedly repair them, \toolname{} recognizes and preserves valid failure-inducing tests, thereby enabling the generation of non-regression tests that can reveal real bugs.
To this end, we conduct a manual evaluation of 167 suspected fault-revealing test cases identified by \toolname{} across six projects in RQ1. 

\textbf{Results and Analysis}.
Through a three-round review process conducted by two authors, we find that 154 out of 167 test cases are confirmed as actual bugs, yielding a precision of 92.22\%.
This result highlights the potential of \toolname{} in generating no-regression test cases capable of detecting real-world software bugs.
We further analyze the root causes and impacts of these confirmed bugs to understand the practical value of \toolname{}.
We also examine the independence of the three voting models on the failure cases entering the vote: their pairwise agreement is only fair to moderate (Cohen's $\kappa$ between 0.39 and 0.55), and their individual tendencies to label a failure as a real defect differ considerably (21.26\%, 45.89\%, and 23.97\%), indicating that the three models do not fail in a systematically identical way.

Table~\ref{tab:root_analysis} presents the classifications and distributions of detected bugs.
Regarding root causes, the most common issue is insufficient input validation (58.44\%), typically involving missing boundary checks, type constraints, or logic verification for method parameters.
The next most prevalent causes are unhandled boundary conditions and state inconsistencies, accounting for 20.13\% and 14.29\%, respectively. 
These reflect common oversights in handling special input values and improper internal state management. 
Regarding bug impacts, robustness bugs are predominant (70.13\%), indicating that most failures stem from improper handling of illegal inputs or exceptional cases, affecting system stability and fault tolerance.
Functional bugs constitute 18.83\%, demonstrating that some bugs directly prevented functionality from operating as intended. 
The remaining issues, such as stability (5.19\%), performance (1.30\%), and usability (4.55\%) bugs, are less prevalent but may still degrade user experience or efficiency under certain conditions.

\subsection{Threats to Validity}

\textbf{Data Leakage}.
Since LLMs are pre-trained on extensive public data, our benchmarks may overlap with their training corpus.
To mitigate this risk, we evaluate \toolname{} on a private industrial dataset (UTXXX) containing unseen methods.
Consistent performance across public and internal datasets suggests improvements stem from our advanced framework rather than memorization.

\textbf{Multi-Language Support}.
Our approach is primarily evaluated using Java, raising concerns about its applicability to other languages. To address this, we re-implement our workflow in Python and conduct comparative evaluations against two SOTA Python-specific test generation methods (\textsc{CodaMosa} and \textsc{CoverUp}). The consistent line and branch coverage results indicate that \toolname{} generalizes effectively across programming languages.

\textbf{LLMs Selection}.
Our primary evaluations use GPT-4o, which potentially introduces model-specific biases and limits generalizability. To address this, we conduct parallel experiments with alternative LLMs (i.e., DeepSeek-V3 and Qwen3-30B) under identical conditions. 
Results show consistent relative performance trends across black-box and open-source models, indicating that \toolname{} generalizes well beyond a single LLM.

\textbf{Non-determinism of LLMs}.
Our results are obtained from a single run of each approach due to the substantial monetary and time costs, and the inherent randomness of LLMs may introduce variance in individual metrics.
The consistent trends across six projects, three LLMs, two languages, and an industrial dataset mitigate this threat.

\section{Conclusion}
\label{sec:conclusion}

This paper presents \toolname{}, a novel multi-agent framework for LLM-based unit test generation, which emulates human testing practices with three collaborative agents: a requirement planner, a test generator, and a reviewer agent.
We further equip the agents with external tool APIs and a repository-level knowledge graph, enabling agents to interact with the codebase in a fine-grained and adaptive manner.
Extensive experiments on both Java and Python projects demonstrate that \toolname{} outperforms state-of-the-art LLM-based approaches across various dimensions and achieves substantially higher mutation scores than search-based tools.
Moreover, \toolname{} exhibits strong extensibility across different LLM backbones and practical applicability in industrial scenarios.

\bibliographystyle{IEEEtran}
\bibliography{reference}

\end{document}

%% file: tables/tools.tex
\begin{table*}[t]
\footnotesize
    \centering
    \caption{Implemented Tools Invoked by \toolname{}}
    \label{tab:tools}
    \resizebox{0.8\linewidth}{!}{
    \begin{tabular}{ll}
    \toprule
        \textbf{Tools} & \textbf{Description} \\ 
    \midrule
        \textbf{Graph Manipulation Tools} &   \\ 
    \midrule
        \texttt{add\_method\_summarization} & Adds summarization to the corresponding method.  \\ 
        \texttt{add\_test\_cases} & Adds test cases and test reports to the corresponding focal method.  \\ 
        \texttt{update\_test\_cases} & Updates test cases and test reports for the corresponding focal method.  \\ 
        \texttt{delete\_test\_cases} & Deletes test cases and test reports from the corresponding focal method.  \\ 
    \midrule
        \textbf{Information Retrieval Tools} &   \\ 
    \midrule
        \texttt{find\_variable\_definition} & Finds variable definitions for a given name in the code knowledge graph.  \\ 
        \texttt{find\_method\_definition} & Finds method definitions in the code knowledge graph based on the given name and parameters.  \\ 
        \texttt{find\_class} & Finds the class node for a given class name in the code knowledge graph.  \\ 
        \texttt{find\_method\_calls} & Finds occurrences of a given method being called in the code knowledge graph.  \\ 
        \texttt{find\_method\_usages} & Finds usages of a given method in the code knowledge graph.  \\ 
        \texttt{fuzzy\_search} & Performs a fuzzy search for a given name in the code knowledge graph.  \\ 
        \texttt{search\_similar\_test\_class} & Finds the most relevant class node for a given test class name in the code knowledge graph. \\
    \midrule
        \textbf{Runtime Interaction Tools} &   \\ 
    \midrule
        \texttt{check\_syntax} & Checks whether the syntax of the given test cases is correct. \\
        \texttt{compile\_test\_cases} & Compiles the given test cases and returns True if compilation succeeds. \\
        \texttt{execute\_test\_cases} & Executes the given test cases and returns True if execution succeeds. \\
        \texttt{calculate\_coverage} & Run the given test cases under JaCoCo and report line- and branch-coverage percentages.\\
        \texttt{calculate\_mutation\_score} & Perform mutation analysis with PITest and return the mutation score (= killed / total mutants).\\
    \bottomrule
    \end{tabular}
    }
    \label{tools}
\end{table*}

%% file: tables/dataset.tex
\begin{table}[t]
\footnotesize
\centering
\caption{Details of Evaluation Datasets}
\label{tab:dataset-info}
\resizebox{\linewidth}{!}{%
\begin{tabular}{l|llccc} 
\toprule
\textbf{Name} & \textbf{Project} & \textbf{Abbr.} & \textbf{Version} & \textbf{\#Modules} & \textbf{\#Methods}  \\ 
\midrule
 \multirow{6}{*}{\begin{tabular}[c]{@{}l@{}}Dataset\\ (Java)\end{tabular}} 
 & Commons-Cli  & Cli   & 1.6.0   &     -      & 185        \\
 & Commons-Csv  & Csv   & 1.10.0  &     -      & 172        \\
 & Gson         & Gson  & 2.12.0  &     -      & 480        \\
 & Jfreechart   & Chart & 1.5.5   &     -      & 200        \\
 & Commons-Lang & Lang  & 3.10.0  &     -      & 200        \\
 & Event-Ruler  & Ruler & 1.8.0   &     -      & 217        \\ 
\midrule
\multirow{4}{*}{\begin{tabular}[c]{@{}l@{}}UTXXX\\ (Java)\end{tabular}}
 & Project\_1   & P1    & -       &    -       & 270        \\
 & Project\_2   & P2    & -       &    -       & 91         \\
 & Project\_3   & P3    & -       &    -       & 37         \\
 & Project\_4   & P4    & -       &    -       & 260        \\ 
\midrule
\multirow{3}{*}{\begin{tabular}[c]{@{}l@{}}CM\\ (Python)\end{tabular}}
 & Dataclasses-json &   Data  & 3dc59e0 & 4         & 50         \\
 & Apimd            &   Apimd    & f32841b & 2         & 45         \\
 & Thonny           &   Thonny    & fb389f4 & 3         & 46         \\ 
\bottomrule
\end{tabular}
}
\end{table}

%% file: tables/rq1_performance.tex
\begin{table*}[t]
\footnotesize
\centering
\caption{Effectiveness of test cases generated by \toolname{}}

\label{tab:overall_performance}
\begin{tabular}{cc|ccc|cc|c}
\toprule
\textbf{Approach} & \textbf{Project} & \textbf{Sy. Rate} & \textbf{Co. Rate} & \textbf{Exe. Rate} & \textbf{Li Cov.} & \textbf{Br Cov.} & \textbf{Mut Sc.} \\
\midrule
\multirow{6}{*}{\toolname{}}
& Cli   & 100.00\% & 100.00\% & 100.00\% & 94.05\% & 93.12\% & 76.61\% \\
& Csv   & 100.00\% & 100.00\% & 100.00\% & 97.87\% & 96.99\% & 98.05\% \\
& Gson  & 100.00\% & 97.29\%  & 93.95\%  & 87.07\% & 84.31\% & 80.97\% \\
& Chart & 100.00\% & 100.00\% & 98.50\%  & 97.55\% & 95.38\% & 82.97\% \\
& Lang  & 100.00\% & 100.00\% & 100.00\% & 96.12\% & 94.11\% & 92.93\% \\
& Ruler & 100.00\% & 100.00\% & 98.16\%  & 89.89\% & 87.27\% & 76.52\% \\

\midrule
\multirow{2}{*}{EvoSuite} & \multirow{2}{*}{Total}
  & 100.00\% & \textbf{100.00\%} & 98.07\% & 82.11\% & 80.81\% & 43.59\% \\
 & 
  & (0.00\%) & (↓0.89\%) & (↓0.62\%) & (↑12.46\%) & (↑11.67\%) & (↑91.99\%) \\

\multirow{2}{*}{ChatUniTest} & \multirow{2}{*}{Total}
  & 99.93\%  & 81.29\%  & 66.09\% & 72.40\% & 71.44\% & 51.79\% \\
 &
  & (↑0.07\%) & (↑21.92\%) & (↑47.47\%) & (↑27.54\%) & (↑26.32\%) & (↑61.59\%) \\

\multirow{2}{*}{HITS} & \multirow{2}{*}{Total}
  & 100.00\% & 91.54\%  & 81.71\% & 82.39\% & 81.22\% & 63.14\% \\
 &
  & (0.00\%) & (↑8.27\%) & (↑19.28\%) & (↑12.08\%) & (↑11.11\%) & (↑32.55\%) \\
\midrule
\toolname{}    & Total
  & \textbf{100.00\%} & 99.11\%  & \textbf{97.46\%} & \textbf{92.34\%} & \textbf{90.24\%} & \textbf{83.69\%} \\
\bottomrule

\end{tabular}
\end{table*}

%% file: tables/rq2_ablation.tex
\begin{table}[htbp]
\footnotesize
\centering
\caption{Ablation study of \toolname{}}
\label{tab:ablation-study}
\resizebox{\columnwidth}{!}{
\begin{tabular}{c|cccccc}
\toprule
\textbf{Approach} & \textbf{Sy. Rate} & \textbf{Co. Rate} & \textbf{Exe. Rate} & \textbf{Li Cov.} & \textbf{Br Cov.} & \textbf{Mut Sc.} \\
\midrule
w/o Sim Test Retrieval  & 
100.00\% &
\makecell{98.62\% \\ {\scriptsize($\uparrow$0.50\%)}} &
\makecell{96.84\% \\ {\scriptsize($\uparrow$0.64\%)}} &
\makecell{91.36\% \\ {\scriptsize($\uparrow$1.28\%)}} &
\makecell{89.14\% \\ {\scriptsize($\uparrow$1.45\%)}} &
\makecell{81.52\% \\ {\scriptsize($\uparrow$2.87\%)}} \\
w/o Test Reviewer  &
100.00\% &
\makecell{98.35\% \\ {\scriptsize($\uparrow$0.77\%)}} &
\makecell{94.91\% \\ {\scriptsize($\uparrow$2.69\%)}} &
\makecell{88.53\% \\ {\scriptsize($\uparrow$4.52\%)}} &
\makecell{85.92\% \\ {\scriptsize($\uparrow$5.25\%)}} &
\makecell{77.58\% \\ {\scriptsize($\uparrow$8.09\%)}} \\
w/o Knowledge Graph  &
100.00\% &
\makecell{92.02\% \\ {\scriptsize($\uparrow$7.70\%)}} &
\makecell{86.93\% \\ {\scriptsize($\uparrow$12.11\%)}} &
\makecell{75.65\% \\ {\scriptsize($\uparrow$22.31\%)}} &
\makecell{72.62\% \\ {\scriptsize($\uparrow$24.52\%)}} &
\makecell{66.12\% \\ {\scriptsize($\uparrow$26.83\%)}} \\
\midrule
\toolname{}  & 
\textbf{100.00\%} & \textbf{99.11\%}  & \textbf{97.46\%}  & \textbf{92.34\%} & \textbf{90.24\%} & \textbf{83.69\%} \\
\bottomrule
\end{tabular}
}
\end{table}

%% file: tables/rq3_models.tex
\begin{table}[t]
\centering
\footnotesize
\caption{Performance of \toolname{} with different LLMs}
\label{tab:model_compare}
\resizebox{\linewidth}{!}{%
\begin{tabular}{lcccccc}
\toprule
\textbf{LLM} & \textbf{Sy. Rate} & \textbf{Co. Rate} & \textbf{Exe. Rate} & \textbf{Li Cov.} & \textbf{Br Cov.} & \textbf{Mut Sc.} \\
\midrule
GPT-4o           & 100.00\% & 99.11\% & 97.46\% & 92.34\% & 90.24\% & 83.69\% \\
DeepSeek-V3      & 100.00\% & 99.31\% & 98.62\% & 86.73\% & 84.64\% & 72.11\% \\
Qwen3-30B   & 100.00\% & 94.57\% & 85.63\% & 81.25\% & 78.21\% & 60.44\% \\
\bottomrule
\end{tabular}
}
\end{table}

%% file: tables/disc_industry.tex
\begin{table}[t]
  \centering
  \caption{Comparison Results on Industrial Projects}
  \label{tab:industry_benchmark}
    \begin{tabular}{lccc}
      \toprule
      \textbf{Approach} & \textbf{Li Cov.} & \textbf{Br Cov.} & \textbf{Mut Sc.} \\
      \midrule
      EvoSuite     & 65.60\% & 64.23\% & 36.88\% \\
      ChatUniTest (Qwen3-30B)  & 66.44\% & 65.97\% & 46.22\% \\
      \toolname{} (Qwen3-30B)   & 84.21\% & 82.12\% & 55.38\% \\
      \bottomrule
    \end{tabular}
\end{table}

%% file: tables/disc_readable.tex
\begin{table}[htbp]
\centering
\footnotesize
\caption{Performance of Readability and Usability}
\begin{tabular}{lcccc}
\toprule
\multirow{2}{*}{\textbf{Approach}} & \multicolumn{2}{c}{\textbf{Readability}} & \multicolumn{2}{c}{\textbf{Usability}} \\
\cmidrule(lr){2-3} \cmidrule(lr){4-5}
 & NI & CL & AQ & AE \\
\midrule
TestAgent & \textbf{2.70} & 2.72 & \textbf{2.95} & \textbf{2.71} \\
EvoSuite & 1.78 & 2.27 & 2.07 & 2.05 \\
ChatUniTest & 2.55 & \textbf{2.80} & 2.76 & 2.65 \\
HITS & 2.38 & 2.45 & 2.49 & 2.33 \\
\bottomrule
\end{tabular}
\label{tab:readability_results}
\end{table}

%% file: tables/disc_python_result.tex
\begin{table*}[htbp]
\footnotesize
\centering
\caption{Comparison Results on Python Projects}
\label{tab:coverage-comparison}
\begin{tabular}{llcccccc}
\toprule
\multirow{2}{*}{\textbf{Project}} & \multirow{2}{*}{\textbf{Module}} 
& \multicolumn{2}{c}{\textbf{\textsc{CodaMosa}}} 
& \multicolumn{2}{c}{\textbf{\textsc{CoverUp}}} 
& \multicolumn{2}{c}{\textbf{\toolname{}}} \\
\cmidrule(lr){3-4} \cmidrule(lr){5-6} \cmidrule(lr){7-8}
& & Li Cov. & Br Cov. & Li Cov. & Br Cov. & Li Cov. & Br Cov. \\
\midrule
\multirow{4}{*}{Data} 
& cfg        & 96.36\% & 87.50\% & \textbf{100.00\%} & \textbf{100.00\%} & \textbf{100.00\%} & 95.45\% \\
& undefined  & 44.26\% & 22.03\% & 86.34\%  & 50.00\%  & \textbf{90.71\%}  & \textbf{58.33\%} \\
& core       & 18.06\% & 6.78\%  & 60.79\%  & 51.35\%  & \textbf{88.11\%}  & \textbf{82.88\%} \\
& mm         & 47.72\% & 14.48\% & 72.55\%  & 53.25\%  & \textbf{80.00\%}  & \textbf{63.64\%} \\
\midrule
\multirow{2}{*}{Apimd} 
& loader     & 69.27\% & 50.64\% & 35.78\%  & 5.26\%   & \textbf{86.24\%}  & \textbf{84.21\%} \\
& parser     & 52.89\% & 36.86\% & 48.66\%  & 35.90\%  & \textbf{84.82\%}  & \textbf{70.94\%} \\
\midrule
\multirow{3}{*}{Thonny} 
& jedi\_utils      & 71.11\% & 44.23\% & 73.33\%  & 45.00\%  & \textbf{93.33\%}  & \textbf{95.00\%} \\
& pgzero & 45.45\% & 60.00\% & \textbf{100.00\%} & \textbf{100.00\%} & \textbf{100.00\%} & \textbf{100.00\%} \\
& roughparse       & 40.10\% & 20.46\% & 52.94\%  & 30.05\%  & \textbf{76.47\%}  & \textbf{59.59\%} \\
\midrule
Total & -        & 53.91\% & 38.11\% & 70.04\% & 52.31\% & \textbf{88.85\%} & \textbf{78.89\%} \\
\bottomrule
\end{tabular}
\end{table*}

%% file: tables/disc_bug_detect_desc.tex
\begin{table*}[htbp]
\footnotesize
  \centering
  \caption{Taxonomy of Root Causes and Bug Impacts}
  \label{tab:root_analysis}
  \resizebox{0.95\textwidth}{!}{%
  \begin{tabular}{llrl}
    \toprule 
    \multicolumn{1}{l}{\textbf{Type}} & \textbf{Category} & \textbf{Ratio} & \textbf{Description} \\
    \midrule
    \multirow{5}{*}{\textbf{Root Causes}} & Insufficient input validation & 90(58.44\%) & Failure to check boundary conditions, or business-logic validity of method parameters. \\
          & Unaddressed boundary conditions & 31(20.13\%) & Exceptional values (e.g., 0, extreme values, empty collections, null) are ignored. \\
          & Loop/iteration errors & 3(1.95\%) & Incorrect or missing termination conditions lead to infinite loops or improper traversal. \\
          & Missing exception handling & 8(5.19\%) & Potential exceptions are neither caught nor correctly propagated. \\
          & Inconsistent state & 22(14.29\%) & Internal object state is not properly reset or initialized, resulting in data corruption. \\
    \midrule
    \multirow{5}{*}{\textbf{Bug Impacts}} & Functional bug & 29(18.83\%) & The functionality does not behave as intended. \\
          & Stability bug & 8(5.19\%) & Causes system crashes, deadlocks, infinite loops, or similar failures. \\
          & Robustness bug & 108(70.13\%) & Improper handling of invalid inputs results in exceptions. \\
          & Performance bug & 2(1.30\%) & Code executes but with low efficiency, leading to resource wastage. \\
          & Usability bug & 7(4.55\%) & Error messages are uninformative, leaving users unaware of the cause or remedy. \\
    \bottomrule
  \end{tabular}
  }
\end{table*}